\documentclass[aps,prl,amsmath,amssymb,preprint,superscriptaddress,showpacs,showkeys]{revtex4-1}
\usepackage[pdftex]{graphicx}
\usepackage{dcolumn, bm, hyphenat}
\usepackage[utf8]{inputenc}
\usepackage[T1]{fontenc}
\usepackage[english]{babel}
\usepackage[mathlines]{lineno}
\usepackage{blindtext}
\usepackage{color}

\newcommand{\ADominance}{{\textit{i}}}
\newcommand{\Consensus}{{\textit{ii}}}
\newcommand{\BDominance}{{\textit{iii}}}
\newcommand{\Polarization}{{\textit{iv}}}
\newcommand{\cotan}{\mathrm{cotan}}
\begin{document}
\title[Consensus and Polarisation in Competing Complex Contagion Processes]{Consensus and Polarisation in Competing\\Complex Contagion Processes}
\author{Vítor V. Vasconcelos}
\email{vvldv@princeton.edu, fpinheiro@novaims.unl.pt}
\affiliation{Department of Ecology and Evolutionary Biology, Princeton University, Princeton (NJ), USA}
\author{Simon A. Levin}
\affiliation{Department of Ecology and Evolutionary Biology, Princeton University, Princeton (NJ), USA}
\author{Flávio L. Pinheiro}
\email{vvldv@princeton.edu, fpinheiro@novaims.unl.pt}
\affiliation{Nova Information Management School (NOVA IMS), Universidade Nova de Lisboa, Lisboa, Portugal}
\affiliation{The MIT Media Lab – Massachusetts Institute of Technology, Cambridge (MA), USA}
\date{\today}
\begin{abstract}
The rate of adoption of new information depends on reinforcement from multiple sources in a way that often cannot be described by simple contagion processes. In such cases, contagion is said to be complex. Complex contagion happens in the diffusion of human behaviours, innovations, and knowledge. Based on that evidence, we propose a model that considers multiple, potentially asymmetric, and competing contagion processes and analyse its respective population-wide dynamics, bringing together ideas from complex contagion, opinion dynamics, evolutionary game theory, and language competition by shifting the focus from individuals to the properties of the diffusing processes. We show that our model spans a dynamical space in which the population exhibits patterns of consensus, dominance, and, importantly, different types of polarisation, a more diverse dynamical environment that contrasts with single simple contagion processes. We show how these patterns emerge and how different population structures modify them through a natural development of spatial correlations: structured interactions increase the range of the dominance regime by reducing that of dynamic polarisation, tight modular structures can generate structural polarisation, depending on the interplay between fundamental properties of the processes and the modularity of the interaction network. 
\end{abstract}
\maketitle

\section*{Introduction}
The study of how information -- opinions, diseases, innovations, norms, attitudes, habits, or behaviours -- spreads throughout social systems has occupied the physical and social sciences for decades \cite{broadbent1957percolation, callaway2000network, xia2011opinion, bornholdt2011emergence, acemoglu2011opinion, valdano2018epidemic, centola2018behavior, guilbeault2018complex}. In that context, the propagation of information has traditionally been assumed to happen through simple-contagion -- a contact process in which information spreads through pairwise interactions \cite{weidlich1971statistical, valente1996network, castellano2009statistical, mobilia2003does, durrett2005can, Sood:2005fr, sood2008voter, masuda2010heterogeneous, conover2011political, banisch2012agent, Pinheiro:2014hv}. However, recent empirical evidence suggests that different types of information spread differently \cite{karsai2014complex, guevara2016research, gao2017collective, hidalgo2018principle, pinheiro2018shooting,Alshamsi:2018tt,weng2012competition,myers2012clash}. In particular, the acquisition of information that is either \textit{risky}, \textit{controversial}, or \textit{costly} seems to require reinforcement from multiple contact sources \cite{centola2018behavior,guilbeault2018complex}. Contrary to simple-contagion, these processes result in propagation impediments to, and through, isolated regions of social networks \cite{nematzadeh2014optimal} and were coined as complex contagion. Its implementation typically occurs in the  context of cascading phenomena \cite{campbell2013complex}, where each individual is activated with some property if a set fraction of neighbours already has that property. Several tools and different perspectives have addressed complex contagion in a context of competition between opposing information elements, but little has been done in merging it to the literature of population dynamics \cite{galam1991towards,galam2008sociophysics,fennell2019multistate,klimek2019fashion,chang2018co, vazquez2010agent}.

A related literature regards that of opinion formation, for which the Voter Model (VM) is a prototypical example. The unpublished work by Durrett and Levin, referenced in \cite{Ehrlich2005Norms}, showcases the competition of complex-contagion processes by introducing the threshold-VM, where two processes with fixed thresholds like the one described above propagate simultaneously. However, empirical work suggests that individuals' response -- given all underlying processes of decision making -- does not result in fixed thresholds but provides support for a range of smooth responses to the neighbourhood configuration \cite{karsai2014complex, guevara2016research, hidalgo2018principle, pinheiro2018shooting, Alshamsi:2018tt}. A body of literature that comes close to a smooth response of individuals to their surroundings is that which is concerned with mean-field approximations of the $q$-VM \cite{Przybyla2011,moretti2013mean}. There, individuals change their state if a neighbourhood of size $q$ has, unanimously, the opposing state. Even though this is a version of the threshold-VM with a unanimous threshold, the mean-field approximations provide a description that is not discrete, even if the underlying model is. Other works \cite{wieland2013asymmetric,min2018competing} break the symmetry between the propagating processes by considering either competition between or coexistence of simple- and complex-contagion processes. An important generalisation of these processes was developed independently by Molofsky et al. \cite{molofsky1999local}, in a model considering competition between two species happening in space (lattice): depending on the number of individuals of each species, say $n$ of species 1, there is a probability vector $p_n$ that a target site becomes a 1. These effective probabilities are necessarily a combination of the processes happening between the two specific species. The model we propose is also general and adequate to be extracted from data for particular processes but explicitly characterises each process instead of their combination \cite{abrams2003linguistics}. Here, we provide a characterisation of the spread of competing information under generalised complex contagion processes. 

We want to understand the effect for the outcome on collective behaviour of the interaction between the propagation of opposing information (in its broad sense) with intrinsically different contagion properties. What is the impact of such interaction in terms of social influence and its dependence on the social structure? On a population with competing processes, we first need to characterise realistic underlying processes and describe their dynamics, simplistically interpolating different behaviours. For adherence to reality, the underlying process must be adaptable and, preferably, built in a way that is measurable as a property of the type of process involved. Below, we provide a characterisation of the spread of competing information under complex contagion, describing each piece of information as a non-linear non-deterministic complex contagion. We show the emergence of different patterns for different types of networks and delineate their macroscopic properties. By doing so, we make explicit the relationship between two seemingly different dynamics: opinion formation and complex contagion.

\section*{Model}
\begin{figure*}[!t]
    \centering
    \includegraphics[width=1.0\textwidth]{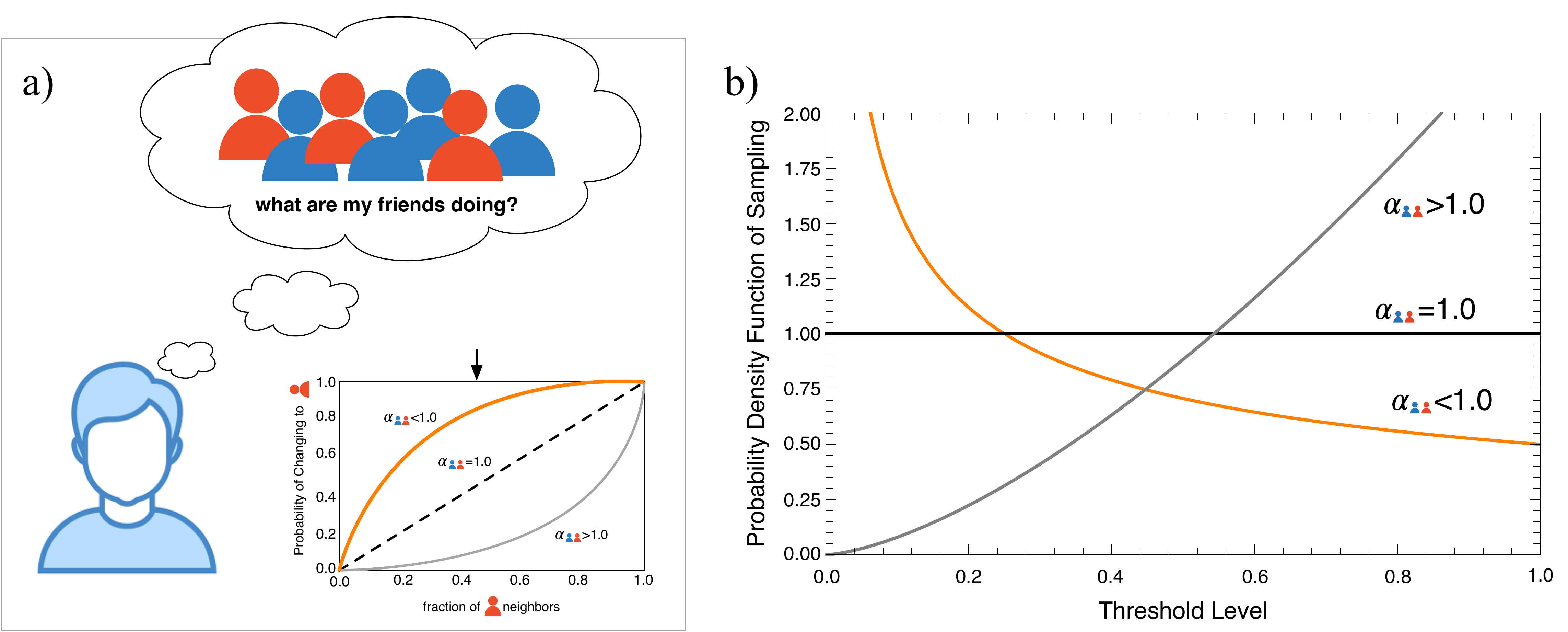}
    \caption{\textbf{a}) Graphical depiction of the complex contagion process under study. Individuals revise their state taking into consideration those of their friends, which, jointly with the complexity of the diffusing information, define the likelihood of an update (see Equation~\eqref{eq:one}). \textbf{b}) A popular approach in the literature is to consider threshold models, in which the probability of updating to a new state is one if the number of friends in such state is above a specified threshold, being zero otherwise. Here we show that our model corresponds to considering a mean-field description of a population with a distribution of thresholds and holds similar dynamical properties. For instance, lower complexities are dynamically equivalent to scenarios with lower thresholds, while high complexities return a similar outcome as expected from high threshold processes.
    }
\label{figure1new}
\end{figure*}

Our processes progress in a finite population of $Z$ individuals who hold alternatively one of two opinions, say $A$ or $B$. For convenience, we discuss our results in light of a population holding different opinions. It is, however, important to stress that our model is more general and can describe any non-overlapping information that can diffuse trough a population. At each moment, there are $k$ individuals with opinion $A$ and $Z-k$ with opinion $B$. The number of contacts an individual $i$ has defines its degree, $z_i$. Individuals revise their opinion by taking into consideration the opinion composition in their neighbourhood. We assume that such events are taken unilaterally by the individuals and that the likelihood that individuals successfully update their opinion is reinforced by multiple contact sources, the definition of complex contagion \cite{centola2018behavior}. Formally, an individual $i$ with opinion $X$ and $n_i^Y$ contacts with opinion $Y \neq X$ changes to opinion $Y$ with some probability function. Let us consider the following function
\begin{equation}
\label{eq:one}
    p_i^{X \rightarrow Y} = \bigg (\frac{n_i^Y}{z_i} \bigg)^{\alpha_{XY}},
\end{equation}
where $X$ and $Y\neq X$ can take the values $A$ or $B$, see Figure~\ref{figure1new}. The probabilities $p_i^{X \rightarrow X}$ are given by the reciprocals of $p_i^{X \rightarrow Y}$. Equation \ref{eq:one} allows us to interpolate between scenarios where opinions require few and many reinforcement sources to propagate (notice that with $\alpha_{XY}=1$ the probability of changing strategy is that of a voter model). We say that $\alpha_{XY}$ accounts for the complexity of opinion $Y$ when learned by an individual that holds opinion $X$. Simpler opinions require less reinforcement from peers than complex ones. When $\alpha_{XY} \neq \alpha_{YX}$, we say the population evolves under asymmetric complexities, while symmetric scenarios happen whenever $\alpha_{XY}=\alpha_{YX}$. In the latter case the dynamics are not affected by an interchange of opinions. 

This notion of complexity of acquisition can be better understood in the light of the parallel with the thresholds models we discuss in the introduction \cite{o2015mathematical, granovetter1978threshold}. Our approach directly maps to a scenario in which individuals are not restricted to having a fixed threshold. As mentioned, a fractional threshold implies that there is a well-defined threshold of neighbours above which an individual adopts new information and below which does not. Dynamically, such a definition results in a deterministic process that either percolates or becomes contained to a few elements of the system \cite{watts2002simple,kempe2003maximizing,watts2007influentials,roukny2013default}. Intuitively, higher thresholds correspond to higher complexities, information that is more difficult to acquire, whereas lower thresholds correspond to lower complexities. One can consider that this threshold is not fixed for all individuals but, instead, as a non-homogeneous distribution. Figure \ref{figure1new}b shows different examples of probability distributions from which individuals can sample their threshold. Let us say individuals change their opinion if the fraction of neighbours with that opinion is greater than a threshold, $0\le M\le1$. That threshold is not fixed but has a probability distribution $d(M)$ given, e.g., by $d(M)\propto M^{\alpha_{AB}-1}$, which can also be seen as a mean-field description of heterogeneous thresholds throughout the population \cite{karsai2016local,unicomb2018threshold}. In this convenient case, $\alpha_{AB}=1$ provides no bias towards high or low thresholds, $\alpha_{AB}<1$ has a bias towards low threshold (representing a simpler adoption process), and $\alpha_{AB}>1$ has a bias towards high threshold (representing a more complex adoption process). Then, the probability that an individual changes from opinion $A$ to $B$ is the probability that it samples a threshold smaller than the fraction of individuals of opinion $B$ in its neighbourhood, $x_i^B$, i.e., $p_i^{A\rightarrow B}=\int_0^{{x_i^B}}{d(M)dM}=({x_i^B})^{\alpha_{AB}}$, corresponding to Equation(\ref{eq:one}) with $x_i^B \equiv n_i^B/z_i$. Different distributions will result in different shapes of $p_i^{A \rightarrow B}$. We provide an analysis of a general $p_i^{X \rightarrow Y}$ in the Electronic Supplementary Material (ESM). 

Our model contains the key elements to study complex contagion of competing non-overlapping processes, though, as stated above, alternative forms could be used \cite{molofsky1999local}. In particular, we will show that even being non-deterministic, it allows for ideas to be contained (for arbitrarily long times) in a subset of the population \cite{centola2007complex}, both in dynamic and structural ways.

\section*{Results and Discussion}
\subsection*{Complex Contagion in Well-Mixed Populations}
\begin{figure}[!t]
    \centering
    \includegraphics[width=1.0\linewidth]{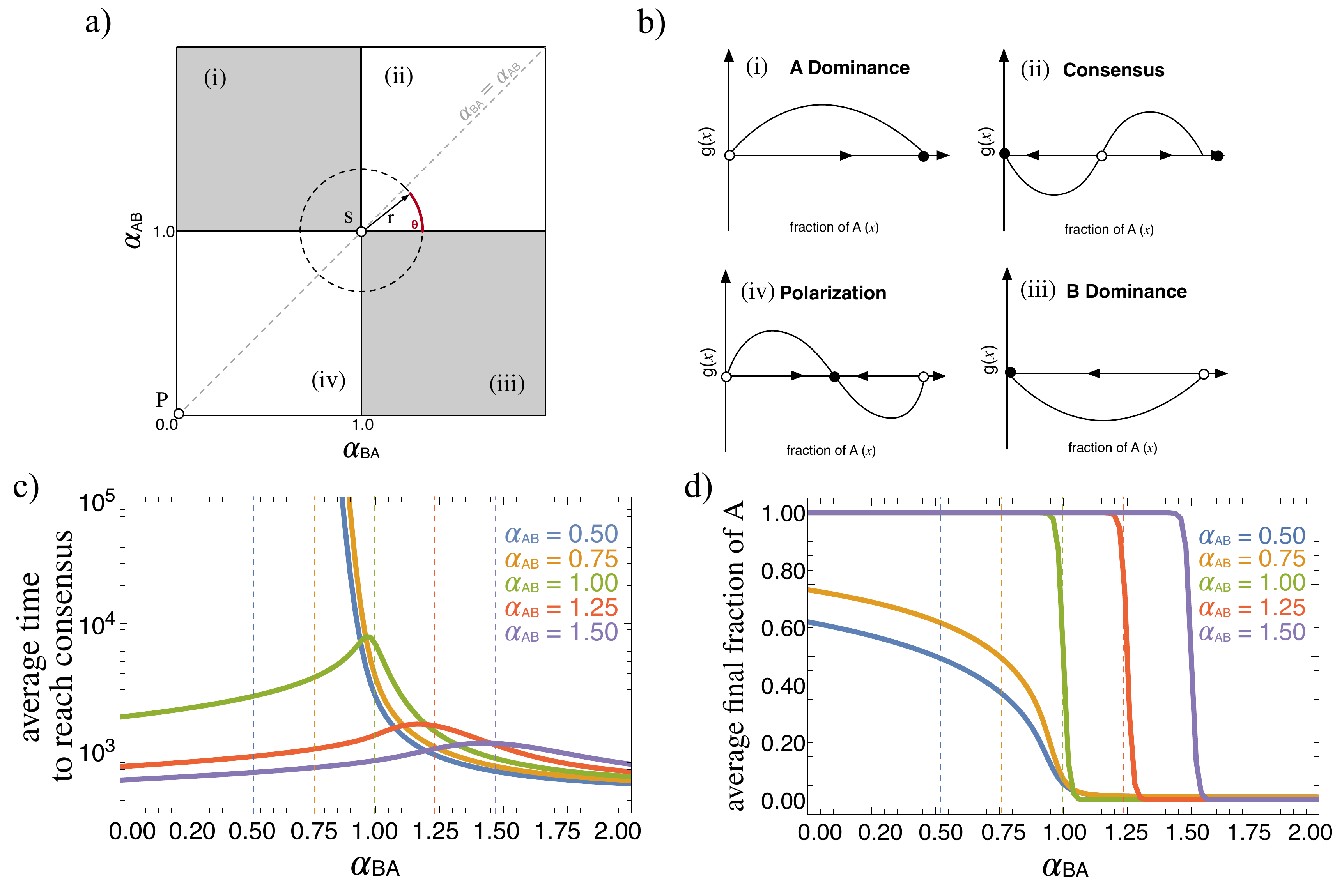}
    \caption{Opinion dynamics under complex contagion in well-mixed populations. 
        \textbf{a}) the four different dynamical regions mapping into the parameter space ($\alpha_{AB}$,$\alpha_{BA}$). Dashed circle represents the polar reparametrisation of the complexity parameters given by $\alpha_{BA}=1+r\cos{\theta}$ and $\alpha_{AB}=1+r\sin{\theta}$ which we use in the remaining of the manuscript to classify the four regions according to $\theta$. Notice that $\gamma=\cotan(\theta)$, which defines the position of the internal fixed point in Equation(\ref{eq:threeone}), is independent of $r$.
        \textbf{b}) the four possible dynamical patterns (\textit{i.e.}, shapes of $g(x)$) obtained, which correspond to Dominance of $B$ (\BDominance) or $A$ (\ADominance), Polarisation (\Polarization), and Consensus (\Consensus).
        \textbf{c}) Average fixation times to reach consensus, when starting from a configuration with equal prevalence of both opinions. \textbf{d}) Average fraction of $A$s in the equilibrium. In \textbf{c}) and \textbf{d}) vertical dashed lines indicate $\alpha_{AB} = \alpha_{BA}$ and populations have $100$ individuals.
    }
\label{figure1}
\end{figure}

We model propagation by means of a network of social interactions, where nodes correspond to individuals and links to social interactions between pairs of individuals. A limiting case happens when we consider well-mixed populations ($z_i=Z-1$), those in which individuals can interact with the entire population without constraints. In such a case, all possible configurations of the system can be defined by $k$, the number of individuals with opinion $A$, and assessing the transition probabilities between configurations allows the full description of the dynamics. This scenario is essential for setting a baseline in which all individuals in the network are identical and there are no spacial correlations. It will be crucial for characterising and understanding the dynamics in different structures. For this, we can use standard procedures to describe the dynamics of such a system. Assuming the probability that two changes of opinions occur in a small time interval, $\tau$, to be $O(\tau^\beta)$, with $\beta>1$, we can reduce the problem to a one-step process \cite{van1992stochastic} and simply compute the probabilities of increasing or decreasing $k$ by one ($0 \leq k \leq Z$), respectively,
\begin{equation}
\label{eq:two}
         T^+_k = \frac{Z-k}{Z}p^{B\rightarrow A} \text{ and }
         T^-_k = \frac{k}{Z}p^{A\rightarrow B}.
\end{equation}

The rate of change in the average abundance of individuals with opinion $A$ is given by the so-called gradient of selection, $g(x)$ \cite{traulsen2005coevolutionary,pacheco2014climate}, where $x \equiv k/Z$. The gradient of selection is equivalent to the drift term in the Fokker-Planck equation describing the probability distribution of the stochastic processes \cite{van1992stochastic} and thus can be used to accurately characterise properties of finite population distributions \cite{vasconcelos2017stochastic}. In the limit of very large populations, $Z \rightarrow \infty$, the dynamics of the fraction of individuals becomes deterministic and can be described by a non-linear differential equation of the form
\begin{equation}
\label{eq:three}
    \dot{x} \equiv g(x) = x(1-x)(x^{\alpha_{BA}-1}-(1-x)^{\alpha_{AB}-1}),
\end{equation}
which, for $\alpha_{XY} \neq 1$, has two trivial solutions --  at $x=0$ and $x=1$ -- and an additional internal fixed point that can be inspected by solving the transcendental equation 
\begin{equation}
\label{eq:threeone}
    1-x=x^\gamma,
\end{equation}
where $\gamma = (\alpha_{BA}-1)/(\alpha_{AB}-1)$. A detailed derivation of Equation~\eqref{eq:three} and Equation~\eqref{eq:threeone} can be found in the ESM.

Equation \eqref{eq:three} holds a similar form to the Replicator Equation from Evolutionary Game Theory \cite{cressman2003evolutionary}, where $x^{\alpha_{BA}-1}$ and $(1-x)^{\alpha_{AB}-1}$ play the roles of the fitness of individuals with opinion $A$ and $B$, respectively. Indeed, the dynamical patterns of opinion Dominance, Polarisation, and Consensus derived from Equation~\eqref{eq:three} are identical to the Prisoner’s Dilemma, Stag Hunt, and Snowdrift Game \cite{sigmund1999evolutionary, santos2006evolutionary} so often studied in that literature. This result provides another interpretation of competitive complex contagion processes: while here individuals change opinions unilaterally, this process is equivalent to a fitness-driven contact process. Importantly, Equation~\eqref{eq:three} also makes clear that, when implementing complex contagion in a competition context, one ends up in a formulation that is equivalent to previous models of opinion formation. Additionally, when one of the complexities is infinite, making opinion adoption irreversible, we recover the dynamics of a single cascading contagion.

Figure~\ref{figure1}a shows how the different dynamical patterns map into the $\alpha_{AB} \times \alpha_{BA}$ domain, while Figure~\ref{figure1}b illustrates the different shapes of $g(x)$ that defining the dynamics in each region. In region (\Consensus), $g(x)$ is characterised by an unstable internal fixed point leading to a coordination-like dynamics towards a consensus, which depends only on the initial abundance of opinions. In region (\Polarization), $g(x)$ has an internal stable fixed-point that results in the polarisation of opinions, which is distinguished by the sustained prevalence of both opinions. In both cases, the specific location of the internal fixed point depends only on the relationship between the complexities of both opinions. In regions (\ADominance) and (\BDominance), $g(x)$ does not have any internal fixed point, and one of the two opinions will invariably dominate the population. We represent two additional special points, $S$ and $P$. In $S$, $\alpha_{AB} = \alpha_{BA} = 1$, every possible configuration of the system corresponds to a fixed point, and finite populations evolve under neutral drift. For $P$, $\alpha_{AB} = \alpha_{BA} = 0$, the dynamics on finite populations reduce to an Ornstein-Uhlenbeck process, with linear drift and constant diffusion.

While Equation\eqref{eq:three} describes large populations, finite populations are characterised by stochastic effects that perturb the system whenever diffusion is non-zero. A particularly interesting measure in finite populations is the time ($\tau_{k_0}$) the population takes to reach a consensus when starting from configuration $k_0$. Since, for $\alpha_{XY}>0$, the system has two absorbing states, at $k=0$ and $k=Z$, it constitutes an Absorbing Markov Chain. Thus, the time to consensus (fixation) starting from configuration $k$ can be formally computed as \cite{traulsen2009stochastic}
\begin{subequations}
\begin{align}
    &\tau_k = -\tau_1 \sum_{j=k}^{Z-1}\prod_{m=1}^{j}\gamma_m + \sum_{j=k}^{Z-1}\sum_{l=1}^{j}\frac{1}{T^+_l}\prod_{m=l+1}^{j}\gamma_m\\
    &\tau_1 = \frac{1}{1+\sum_{j=1}^{Z-1}\prod_{m=1}^{j}\gamma_m}\sum_{j=1}^{Z-1}\sum_{l=1}^{j}\frac{1}{T^+_l}\prod_{m=l+1}^{j}\gamma_m
\end{align}
\end{subequations}
where $\gamma_m = T_m^-/T_m^+$.

Although finite populations always reach a consensus, the time required to do so can be extremely long, in particular for region (\Polarization). Indeed, for that region, even for small populations of $100$ individuals, the time to consensus is of the order of $10^{13}$ generations, making it more likely to find the population in a polarised state, as we would expect from the existence of a stable fixed point in the corresponding deterministic limit. Figure~\ref{figure1}c shows the time required for reaching consensus starting from a perfect mix of opinions (50-50) for different combinations of $\alpha_{AB}$ and $\alpha_{BA}$. Figure~\ref{figure1}d shows, under the same conditions, the expected evolutionary outcome after $10^6$ generations. In ESM, we show that the fixation time has a non-monotonous relationship with complexity by varying $r$ at a fixed angle. Indeed, assuming fairly distributed initial conditions, for high adoption-complexity ideas (corresponding to region (\Consensus), i.e., $\theta\in(0,\pi/2)$), there is a complexity value that minimises the time to fixation: for small $r$ the dynamics are close to a random walk, as described for point S; as $r$ increases, selection of opinions through coordination increases the speed of conversion, to one or either side; finally, for large complexity, the adoption rates decrease again, as it becomes harder to find neighbourhoods containing enough reinforcement for either opinion.

\subsection*{Complex Contagion Processes in Complex Networks} 
\begin{figure*}[!t]
    \centering
    \includegraphics[width=1.00\textwidth]{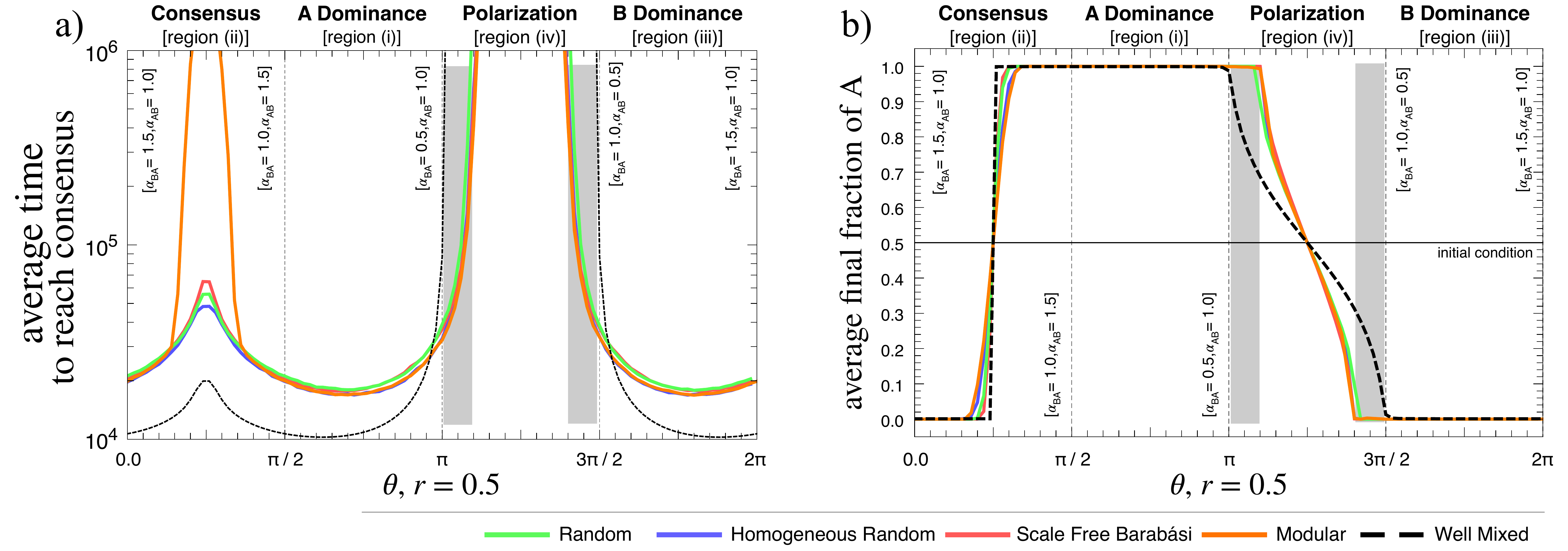}
    \caption{Opinion dynamic under complex contagion in structured populations.
    Panel \textbf{a}) shows the fixation times, measured in Monte Carlo steps, for different population structures (green, red, orange, and blue curves) and well-mixed (dashed black lines), starting from a 50-50 distribution of As and Bs. We measure time in Monte Carlo steps. Panel \textbf{b}) shows the average final fraction of individuals with opinion $A$ in the population for different population structures (green, red,  orange, and blue curves) and well-mixed (dashed black lines). For a convenient comparison between all types of population structures, shaded areas indicate the mismatch between well-mixed and structured populations in the Polarisation region (\Polarization). The complexity parameters have been reparameterised as $\alpha_{BA} = 1 + r \cos\theta$ and $\alpha_{AB} = 1 + r \sin\theta$, with $r =1/2$. Notice that $\gamma=\cotan~\theta$, which defines the position of the internal fixed point in Equation(\ref{eq:threeone}), is independent of $r$. Other parameters: $Z = 10^3$ and average degree is $4$.
    }
\label{figure2}
\end{figure*}

Most application cases do not fall into the category of the previous section. In fact, social systems, but also propagation of properties resulting from aggregated processes in abstract networks, are marked by peer influence, which induces spatial correlations between individuals, or nodes, in their interacting network. We turn our attention to the case in which individuals are only able to interact with a small subset of the population ($z_i < Z$). We focus on five network topologies with a fixed average degree: Lattice with periodic boundaries, Homogeneous Random \cite{santos2005epidemic}; Erdős–Rényi Random \cite{erdos1960evolution}; Scale Free Barabási-Albert \cite{albert2002statistical}; and a modular network that is comprised of two Scale Free Barabási-Albert networks where 4\% of the nodes have internetwork uniformly random connections. A detailed description of how these networks were generated is in the ESM.

Figure~\ref{figure2} compares the results of opinion dynamics under complex contagion in different populations structures along the four dynamical regions we previously identified: Consensus, $A$ Dominance, Polarisation, and $B$ Dominance. Generally, the dynamical patterns observed in structured populations follow what we previously observed in well-mixed populations. However, there are two eye-catching differences. Population structure significantly shortens the domain where Polarisation is observed (region \Polarization), increasing Dominance. The mismatch is highlighted in Fig~\ref{figure2}b with a grey area. This change in the breath of the polarisation region is a consequence of the spatial correlations introduced by the  structure, which facilitate the dominance of highly populated opinions. However, the same spatial correlations that facilitate dominance slow down the rate at which invasion of opinion happens. Notice how, in Figure~\ref{figure2}a, fixation times, including those for dominance, regions increase in structured populations relative to the well-mixed scenario. That difference is partially due to group-size restrictions and said correlations.  
The second notable difference occurs in the consensus region with the modular networks with a spike in fixation times comparable to the one occurring in region (\Polarization). These networks are distinguished by a funnel effect in which only a few individuals are the bridge between (in our case two) sub-networks or communities. Modular networks introduce a substantial change in the dynamics observed in region (\Consensus). This is due to the fact that now the communities can achieve a distinct consensus and thus generate a polarisation that is driven by an interplay between the network structure and the high complexity of the information, which contrasts with the dynamical polarisation observed in region (\Polarization). This is in fact the strong property of complex contagion, which, to a greater extent, only occurs in this region of complexity. To elucidate this phenomenon we need to better understand, not only the temporal dynamics we described, but the spatial distribution created by the propagation on these structures. 
In mind we keep that, for most connected regions, the dynamical outcomes are identical to but slower than the ones obtained in well-mixed populations and, in particular, the boundaries that divide the regions of behaviours with respect to complexity are well approximated by the mean-field description detailed by Equation~\eqref{eq:three}.

\subsection*{Characterisation of Spatial Correlations} 
\begin{figure*}[!t]
    \centering
    \includegraphics[width=1.00\textwidth]{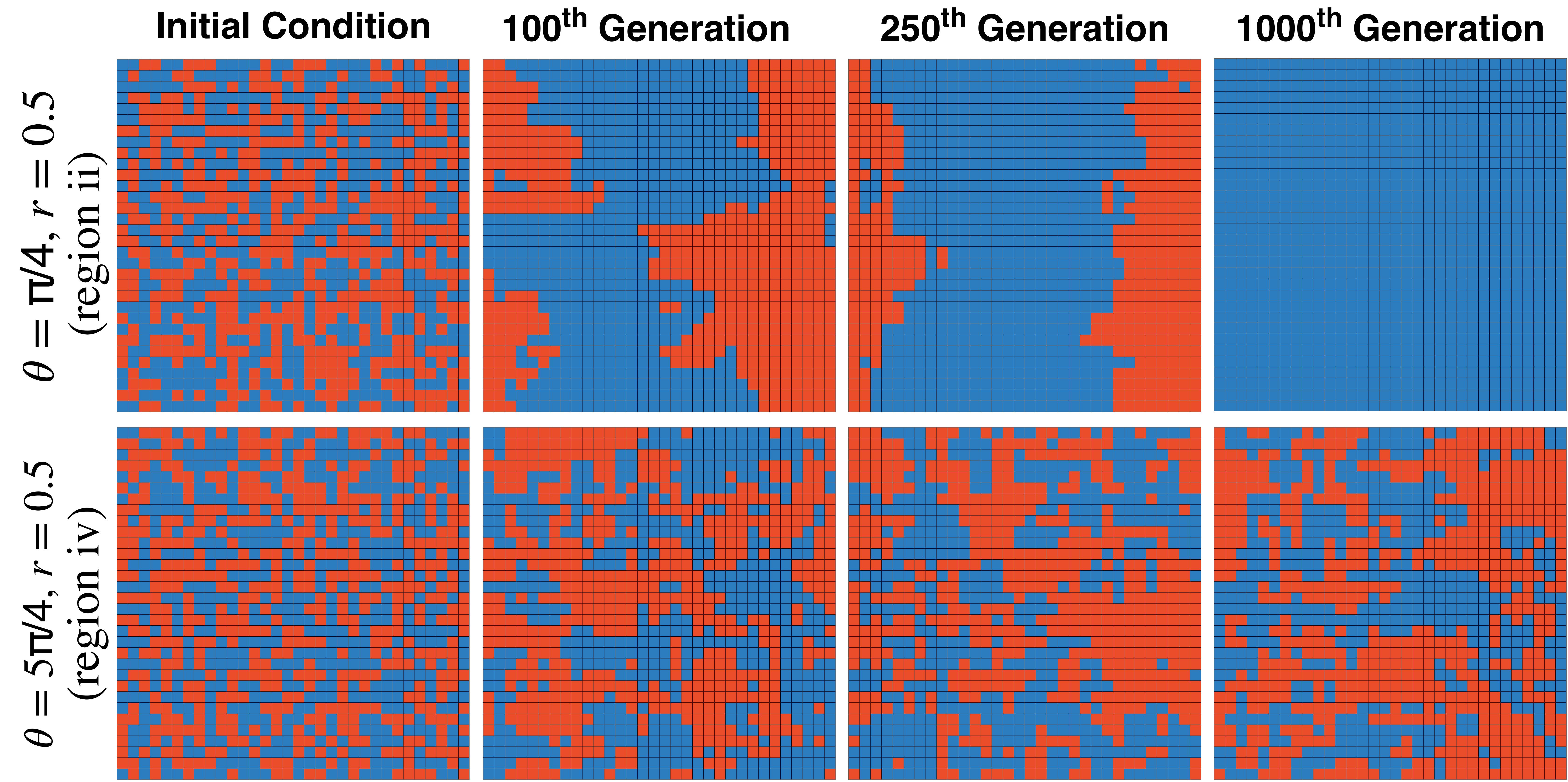}
    \caption{Emergent spatial correlations in the Consensus (top) and Polarisation (bottom) regimes in a square Lattice with $1024$ nodes and degree $4$. Top panels represent a scenario with $\alpha_{AB} = \alpha_{BA} = 1.35$ ($\theta = \pi/4$ and $r = 0.5$) and bottom panels $\alpha_{AB} = \alpha_{BA} = 0.65$ ($\theta = 5\pi/4$ and $r = 0.5$). Under the conditions of the top panels, runs converge to the full dominance of either opinion. Each grid shows a representative snapshot of the emerging spatial arrangements at different times. In each grid, cells are coloured, Blue or Red, according to the opinion of the individual placed in it (symmetric in this case).
    }
\label{figure4}
\end{figure*}

It is well known that information diffusing over a social network generates spatial correlations among traits of individuals and those of their neighbours. Surprisingly, even though our model is suitable to be analysed by pair approximation methods \cite{gleeson2011high,gleeson2013binary}, these correlations tend to extend beyond the dyadic nature of individuals relationships \cite{Pinheiro:2014hv,christakis2007spread}. Likewise, here we show that similar correlations naturally emerge in both the Consensus and Polarisation regimes, although their extension and nature are quite different and dependent on the underlying topological features of the social network.

Figure~\ref{figure4} shows four snapshots at different times of two representative simulations in Lattices, showing the different type of spatial patterns that emerge in the Consensus (top) and Polarisation (bottom) regimes. It is noteworthy to say that the spatial patterns in the top panels are transient, as under the dynamical conditions (high complexity) the population will converge to a consensus. However, for very high complexity, the system would take a long time to move away from its initial condition, a behaviour also found for deterministic threshold models \cite{Ehrlich2005Norms}.

\begin{figure*}[!t]
    \centering
    \includegraphics[width=1.00\textwidth]{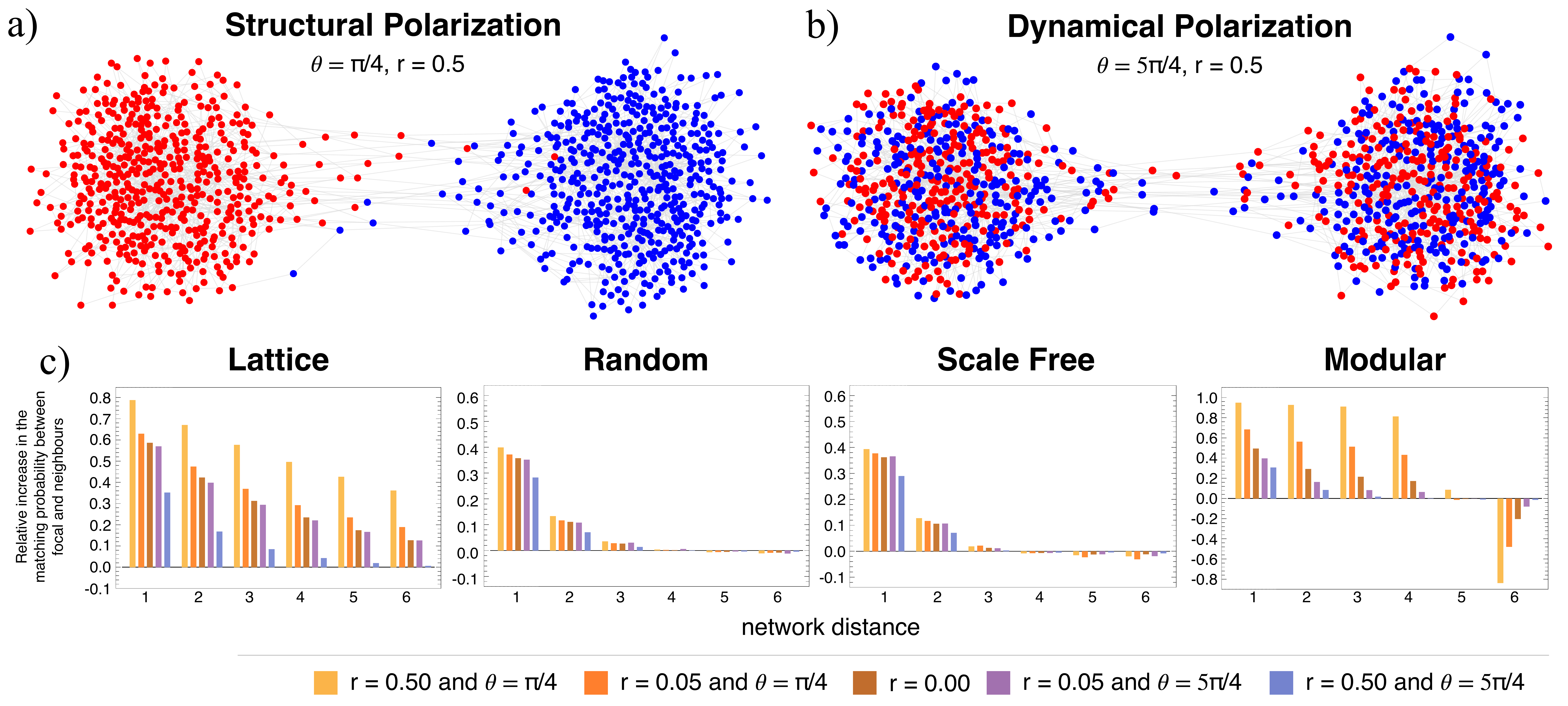}
    \caption{Spatial correlations among individuals with the same opinion.
    Top panels show the possible polarisation outcomes on modular networks, namely  \textbf{a}) shows a dynamical polarisation that results naturally in a scenario with low complexity, while  \textbf{b}) shows a structural polarisation that results from the two modules of the network converge to different opinions.
    \textbf{c}) Shows the relative increase in the probability of the matching of type of a focal node with that of a neighbour at a distance of $n$ links in the network, relative to a random distribution of strategies. These have been computed by sampling spatial distributions at a fixed configuration (50/50) on multiple independent simulations starting from random initial conditions.
    }
    \label{figure5}
\end{figure*}

In the consensus regime, modular structures recover the classical result of complex contagion identified by Centola et al. \cite{centola2007complex}. That is, a structural polarisation that results from each network module reaching a different consensus and then being unable to invade each other (see Figure~\ref{figure5}a). In such a scenario all dynamical activity is restricted to the individuals that lie at the interface between the two clusters. This contrasts with the dynamical polarisation that emerges in the Polarisation regime (see Figure~\ref{figure5}b), which stems from the low complexity of the information being diffused.
Even though those are both scenarios of polarisation, with the population being split between different opinions, they have different statistical properties. In Figure~\ref{figure5}c we compute the increase in the probability that a neighbouring individual at a fixed distance shares the same state as a focal individual relative to random sampling. If this value is positive, a neighbour tends to share the same opinion as a focal individual. If it is negative, a neighbour tends to have the opposite opinion. The distance at which the sign changes correspond to the range of the clustering of opinions. Dynamical polarisation, the one well described by the well-mixed population model (happening in region \Polarization), is characterised by a decay to zero of the relative increase in the probability of matching states as the distance between nodes increases. Structural polarisation, on the other hand, can be identified by a plateau of high increased probability followed by a sharp transition from positive to negative correlation of strategies. For more than two sub-networks, the transition is expected to be smoothed out, as now there are more nodes past the funnel with the same strategy, though the plateau should remain.  

\section*{Conclusions}
In this manuscript we presented a new model of competitive, complex contagion dynamics that contains dynamical patterns of Dominance, Polarisation, and Consensus, depending only on the relative complexity of the diffusing information. These patterns are in many ways equivalent to the ones obtained in other contexts, namely in Evolutionary Game Theory, which deals with problems as diverse as the spin-flips \cite{blume1993statistical}, selection of gut biome \cite{widder2016challenges}, management of common and public goods \cite{tavoni2012survival,vasconcelos2015cooperation}, and socioecological resilience \cite{levin1998resilience}. We show there is an additional a strong equivalence between modelling complex contagion in a competition context and opinion dynamics. Our work raises important questions in terms of feasibility of assessing empirically which mechanisms are at play. Are empirical patterns the result of game-theoretical reasoning of agents that influence strategy adoption or the result of the spreading of information with different levels of complexity? Such an equivalence is in fact enlightening. For instance, the assumption considered here -- that the processes are dependent on the fraction of the neighbouring types -- shows strong invariance to network topology, whereas processes for behaviour change that depend on the total number of neighbours, are known to change the macroscopic nature of the evolutionary dynamics \cite{Pinheiro:2012cn}. This provides a testable way for understanding, in a specific practical situation, which mechanisms are present in decision making by looking for such an invariance. 

Social planners or other network intervenients who aim at understanding the macroscopic behaviour of the population can derive important conclusions from our work. In fact, we show that polarisation can arise both from a modular network structure concerning processes with high adoption complexity, but it can also follow from low acquisition complexity and, in the latter, targeting network structure (e.g.~by making it more modular) or providing particular individuals with incentives towards opinion changing might have a small impact on the final outcome. Alternatively, planners can act in the system by modifying the complexity of what is being spread to improve the chances of getting a dominant behaviour. This, however, can lead to an \textit{arms race} and drive the system to a polarisation trap.

Open questions have been left for future research as, for instance, expanding the model to scenarios that involve more than two competing opinions, the exploration of different functional forms of contagion \cite{chang2018co}, of optimal seeding strategies in competitive scenarios under complex contagion, and the impact of situations in which individuals are able to modify the interaction network \cite{PhysRevE.78.016104,evans2018opinion,chang2018co}. 

\section*{Author Contributions}
V.V.V. and F.L.P. designed the model and analyzed the results. 
V.V.V., S.A.L, and F.L.P. discussed results and wrote the paper.

\section*{Acknowledgements}
The authors are thankful to C\'{e}sar A. Hidalgo, Aamena Alshamsi, Cristian Candia-Castro-Vallejos, and Tarik Roukny for the helpful discussions and insights.

\section*{Data Accessibility}
This manuscript does not use any data.

\section*{Funding Statement}
This work was supported by the Fundação para a Ciência e a Tecnologia (FCT), Portugal through Grants PTDC/MAT-STA/3358/2014 and PTDC/EEI-SII/5081/2014; by the US Defense Advanced Research Project Agency D17AC00005, the National Science Foundation grant GEO-1211972, and the Army Research Office Grant W911NF-18-1-0325; an the Masdar Institute--MIT Cooperative Agreement (USA Ref. 0002/MI/MIT/CP/11/07633/GEN/G/) and the MIT Media Lab Consortia

%
\end{document}